\documentclass{PoS}

\def\ie{{\it i.e.}}
\def\str{{\rm str}}
\def\tr{{\rm tr}}
\def\bq{{\overline q}}

\title{Flavor symmetry breaking in mixed-action QCD}

\ShortTitle{Mixed-action QCD}

\author{Oliver B\"ar\\
        Institute of Physics\\
        Humboldt University, Berlin, Germany\\
        E-mail: \email{obaer@physik.hu-berlin.de}}

\author{\speaker{Maarten Golterman}\\
        Department of Physics and Astronomy\\
        San Francisco State University, San Francisco, CA~94132, USA\\
        E-mail: \email{maarten@stars.sfsu.edu}}

\author{Yigal Shamir\\
        Raymond and Beverly Sackler School of Physics and Astronomy\\
        Tel-Aviv University, Ramat~Aviv, 69978~Israel\\
        E-mail: \email{shamir@post.tau.ac.il}}

\abstract{We study the phase structure of mixed-action QCD with two Wilson sea quarks
and two chiral valence quarks, starting from the chiral lagrangian.
{\it A priori}, the effective theory allows for a rich phase structure, including a
phase with a condensate made of sea and valence quarks.  Because this would
lead to mass eigenstates that are admixtures of sea and valence fields,
pure-sea correlation functions would depend on valence quark masses, in
contradiction with the actual setup of mixed-action simulations.  Using
properties of the chiral Dirac operator, we prove that such a phase does not
occur, and that this leads to bounds on low-energy constants.}

\FullConference{ The XXIX International Symposium on Lattice Field Theory - Lattice 2011\\
July 10-16, 2011\\
Squaw Valley, Lake Tahoe, California}

\begin{document}

\section{Introduction}
In lattice QCD with a mixed action (MAQCD), the discretization of the Dirac operator in the 
sea sector (\ie, the operator from which the fermion determinant is constructed)
is chosen to be different from the discretization of the Dirac operator in the valence
sector (\ie, the operator which is inverted to obtain quark propagators attached to external lines).   

The continuum limit of a mixed-action theory is a partially quenched theory \cite{BGPQ},
because even in the continuum limit the valence and sea quark masses 
$m_v$ and $m_s$ for each
flavor do not have to be equal.  This breaks unitarity already in the continuum limit,
but when the lattice spacing $a\ne 0$, it is not even possible to define equality of
$m_v$ and $m_s$ in a universal way \cite{BRS2}.   This implies, of course, that unitarity
is always broken at $a\ne 0$ in a mixed-action theory.

It is relatively straightforward to check the validity of the mixed-action approach in
weak-coupling perturbation theory.   More non-trivial, non-perturbative information about the validity is obtained by constructing the low-energy effective field theory (EFT) for MAQCD,
mixed-action chiral perturbation theory (MAChPT) \cite{BRS2,BRS1,GIS}.   Chiral perturbation theory (ChPT)
gives access to the phase diagram of the lattice theory, and, as we will see, this 
allows for a non-trivial check of the field-theoretical definition of MAQCD and the
validity of MAChPT.
Here, we investigate the phase structure of MAQCD with two chiral 
(Ginsparg--Wilson \cite{GW}) valence quarks,
and two Wilson 
sea quarks.  For a more detailed account which includes technicalities, we refer to Ref.~\cite{BGS}.
   
\section{MAChPT}
Field-theoretically, MAQCD with two (Wilson) 
sea and two (chiral) valence flavors is a theory with 
six quarks: the two sea quarks, the two valence quarks, and two more ``ghost''
quarks with a Dirac operator identical to that of the valence quarks, but with
opposite statistics, such as to effect a cancellation between valence and ghost
determinants, thus removing both valence and ghost quarks from the sea.

This implies that the non-linear field $\Sigma$ representing the pions in the
EFT is a $6\times 6$ graded matrix
\begin{equation}
\label{Sigma}
\Sigma=\pmatrix{{\rm exp}\left(\frac{2i}{f}\,\phi\right)&{\overline\omega}\cr
\omega&{\rm exp}\left(\frac{2}{f}\,\hat\phi\right)}\ ,
\end{equation}
with $\phi$ and $\hat\phi$ hermitian matrices of size $4\times 4$ and $2\times 2$,
respectively, and $\omega$ and $\overline\omega$ rectangular, Grassmann-values
matrices.  The matrix $\phi$ describes the pions in the sea and valence sectors,
the matrix $\hat\phi$ those made of ghost quarks, while the Grassmann valued 
components describe fermionic pions made out of one ghost, and one sea or valence
quark.  $f$ is the pion decay constant in the chiral
limit.   For an overview of the precise symmetry structure of the theory, which leads
to the unfamiliar way in which $\hat\phi$ appears in the non-linear field $\Sigma$,
see for instance Ref.~\cite{MGLH}.

Since we are only interested in the phase diagram of the theory here, it suffices
to consider the chiral potential, which, in simplified form, and rescaled by the
combination of low-energy constants (LECs) $B_0f^2/4$,
is given by \cite{BRS2}
\begin{equation}
\label{potential}
V=-\str\left(M(\Sigma+\Sigma^{-1})\right)-\frac{c_1}{4}\,\str\left(P_s\Sigma P_s
\Sigma^{-1}\right)-c_2\,\str\left(P_s\Sigma^{-1}P_s\Sigma^{-1}+P_s\Sigma P_s\Sigma
\right)\ ,
\end{equation}
in which $P_s={\rm diag}(1,1,0,0,0,0)$ projects on the sea sector, and $M$ is the quark mass matrix
\begin{equation}
\label{mass}
M={\rm diag}\left(m_s,m_s,m_v,m_v,m_v,m_v\right)\ ,
\end{equation}
where the first two $m_v$ entries correspond to the valence quarks,
and the last two to the ghost quarks.
We consider the ``large cutoff effects'' regime, in which the chiral power counting
is such that $m_s\sim m_v\sim a^2$.   This implies that if we consider terms in the
potential to linear order in the quark masses, we should also consider all terms of
order $a^2$, and the terms proportional to $c_1$ and $c_2$ are such terms.\footnote{Terms linear in $a$ can be absorbed into the quark masses \cite{ShSi}.}
The constant $c_2$ is a LEC special to Wilson
fermions and arises because of chiral symmetry breaking.  Thus, it appears only in the sea sector.   The LEC $c_1$
is special to the mixed-action case, as can be seen from the fact that it disappears
from $V$ when we replace $P_s$ by the unit matrix.  Both $c_1$ and $c_2$ are
lattice artifacts, containing a factor $a^2$.   There are several more terms in the chiral
potential $V$, but those do not affect the discussion of the phase diagram we are
interested in here,
and we therefore omit them in this talk.

Since we will always keep $m_v> 0$ in our discussion below, all the chiral symmetries are explicitly broken (softly in the valence sector), and the full symmetry group $G$ is a generalization of isospin,
\begin{equation}
\label{group}
G=U(2)_{sea}\times U(2|2)_{valence}\ .
\end{equation}
At non-zero lattice spacing there are no symmetries relating the sea and the
valence-ghost sectors, because of the different choices for the sea and valence
Dirac operators \cite{BRS1}.  Within each sector we will always maintain isospin symmetry,
taking the up and down quarks degenerate in each sector.

\section{A puzzle}
Let us expand $V$ around the trivial vacuum in order to obtain the pion masses
at leading order in ChPT (taking $m_{s,v}>0$):
\begin{eqnarray}
\label{pionmasses}
M_{ss}^2&=&2B_0(m_s+4c_2)\ ,\\
M_{vv}^2&=&2B_0m_v\ ,\nonumber\\
M_{sv}^2&=&B_0(m_s+m_v+4c_1+2c_2)\ ,\nonumber
\end{eqnarray}
with $M_{ss}$ the mass of a pion made out of two sea quarks, $M_{vv}$ the 
mass of a pion made out of two valence quarks, and the mixed-pion mass 
$M_{sv}$ of a pion made out of a sea and a valence quark.

The third line of Eq.~(\ref{pionmasses}) suggests that spontaneous symmetry
breaking (SSB) can take place if $2c_1+c_2<0$, because in that case
decreasing the sum $m_s+m_v$ makes $M_{sv}^2$ negative.   This would
drive the theory into a ``mixed phase,'' with a condensate 
$\langle\bq_s\gamma_5 q_v\rangle\ne 0$.  Indeed, if we choose $c_2>0$
(to avoid the complications of an Aoki phase \cite{Aoki} in the sea sector \cite{ShSi}),
but $2c_1+c_2<0$, minimization of the potential $V$ leads to the conclusion
that for $m_s+m_v+4c_1+2c_2<0$ the theory enters a mixed phase.

The mixed condensate would break the symmetry group $G$ down to a diagonal group
in which the remaining sea and valence quark symmetry transformations are the same.  Sea and
valence sectors mix, and this has consequences for the spectrum of the theory.
An interesting example is the two-point function $\langle\pi_{ss}^+(0)\pi_{ss}^-(t)\rangle$,
which is made out of sea pions only, but which in the mixed phase becomes dependent
not only on $m_s$, but also on $m_v$.\footnote{As can be demonstrated explicitly
by a somewhat tedious but straightforward calculation.}

This, however, creates a paradox: by the very construction of MAQCD, this can
never happen!   In an actual simulation, gauge-field configuration depend only
on $m_s$, or, in other words, the dynamics of the theory can only
depend on $m_s$.   In particular, if one considers a correlation function that depends only on sea fields, such as
$\langle\pi_{ss}^+(0)\pi_{ss}^-(t)\rangle$, one performs a simulation in the theory
with only sea quarks; no valence quarks are present anywhere in such a 
computation!\footnote{The paradox persists in finite volume, because
$\langle\pi_{ss}^+(0)\pi_{ss}^-(t)\rangle$ is invariant under integration over
the orientation of $\langle\bq_s\gamma_5 q_v\rangle$ in the mixed phase.}

If this were a true paradox, this would imply that the field-theoretical description
of MAQCD is fatally flawed, and, as a consequence, that standard EFT
techniques cannot be used to interpret results obtained in a mixed-action
simulation.   This state of affairs leads to the following important questions:
\begin{itemize}
\item[1.]  Does MAQCD, in its field-theoretical definition, indeed 
have a mixed phase, \ie, a phase with
$\langle\bq_s\gamma_5 q_v\rangle\ne 0$ (as appears to be predicted by
MAChPT)?
\item[2.]  If not, does MAChPT get it wrong?  
\end{itemize}
In the following section we will see that in fact this dangerous scenario cannot
occur, and that MAChPT is forced, by the underlying theory, to get it right.

\section{Resolving the puzzle}
We begin with a theorem that holds in the underlying theory, MAQCD with
two sea quarks and two valence quarks, invariant under the isospin group
$G$ of Eq.~(\ref{group}).\footnote{In fact, this theorem holds for any 
number of valence quarks.}
  The theorem states that no spontaneous
isospin breaking can occur in the valence sector.  This is an almost direct
consequence of the well-known Vafa--Witten \cite{VW} theorem, that forbids breaking
of vectorlike symmetries in the continuum.  The theorem extends to the valence
sector because even in the lattice theory the valence quarks are chiral; the
Vafa-Witten theorem applies to any type of Ginsparg--Wilson quarks.\footnote{
It can also be shown that the theorem extends to the ghost sector, and,
in particular, that no Grassmann-valued condensates can occur \cite{BGS}.}

An immediate consequence is that a mixed phase cannot occur, because such a phase would
break $U(2)_{sea}\times U(2)_{valence}\to U(2)_{diagonal}$, and thus a
non-zero value of
$\langle\bq_s\gamma_5 q_v\rangle$ would break $U(2)_{valence}$, in 
contradiction with the Vafa--Witten theorem.

This answers the first of the two questions raised in the previous section:
MAQCD cannot have a mixed phase, and the paradox cannot occur.
But, this still leaves open the second question: what about MAChPT?

\section{A mass inequality}
By choosing $m_s>0$ and $m_v>0$ large, we can arrange  that
$M_{ss}^2$, $M_{vv}^2$, and $M_{sv}^2$ are all strictly
positive, so that
no isospin breaking takes place in the EFT.   Then, from the identity
\begin{equation}
\label{start}
\tr\left\langle\left(S_{sd}(x,y)-S_{vd}(x,y)\right)^\dagger
\left(S_{sd}(x,y)-S_{vd}(x,y)\right)\right\rangle\ge 0
\end{equation}
in the underlying theory,
where $S_{si}$ ($S_{vi}$) is the sea (valence) quark propagator for flavor
$i=u,\ d$, and $S_d^\dagger(x,y)=\gamma_5 S_u(x,y)\gamma_5$ for both
sea and valence propagators, it follows that
\begin{equation}
\label{pionineq}
\langle\pi_{ss}^+(x)\pi_{ss}^-(y)\rangle+\langle\pi_{vv}^+(x)\pi_{vv}^-(y)\rangle\ge
\langle\pi_{sv}^+(x)\pi_{vs}^-(y)\rangle+\langle\pi_{vs}^+(x)\pi_{sv}^-(y)\rangle\ .
\end{equation}
Translating this to ChPT, this inequality implies that
\begin{equation}
\label{massineq}
M_{sv}\ge {\rm min}(M_{ss},M_{vv})\ .
\end{equation}
Let us now argue that this inequality restores the validity of MAChPT.  {}From Eq.~(\ref{pionmasses}),
we have that
\begin{equation}
\label{massrelation}
M_{sv}^2=\frac{1}{2}(M_{ss}^2+M_{vv}^2)+2(2c_1-c_2)\ .
\end{equation}
Choosing quark masses such that $M_{ss}=M_{vv}$, the inequality~(\ref{massineq})
implies that $M_{sv}\ge M_{ss}=M_{vv}$, and thus, using Eq.~(\ref{massrelation}),
that
\begin{equation}
\label{LECineq}
2c_1+c_2\ge 2c_1-c_2\ge 0
\end{equation}
(recall that we chose $c_2>0$).  But, now we can turn this around, using the 
fact that the LECs $c_1$ and $c_2$ are {\it independent} of the quark masses
$m_s$ and $m_v$, so that inequality~(\ref{LECineq}), together with the third
equation in Eq.~(\ref{pionmasses}), implies that always,
irrespective of the value of the quark masses, $M_{sv}^2\ge 0$, and no SSB
to a mixed phase can occur, also in MAChPT.   In words, the mass inequality~(\ref{massineq}) restricts the values of the LECs in the EFT such that the EFT is 
forced to faithfully reproduce the phase structure of the underlying theory.
While we demonstrated this here for the simplified potential of Eq.~(\ref{potential}),
and for the case $c_2>0$, it is clear that, in general, the parameters of MAChPT
are restricted such that regions in the phase diagram with a mixed condensate
are excluded, because of the fact that a mixed phase cannot occur in the
underlying theory.

To summarize the situation, the Vafa--Witten theorem restricts the vacuum
expectation value of the non-linear field to the form
\begin{equation}
\label{vev}
\Sigma_{vacuum}=\pmatrix{\Sigma_{ss}&{\bf 0}&{\bf 0}\cr
{\bf 0}&{\bf 1}&{\bf 0}&\cr {\bf 0}&{\bf 0}&{\bf 1}}\ ,
\end{equation}
where $\Sigma_{ss}$ is the $2\times 2$ unitary matrix describing the vacuum
in the sea sector, ${\bf 0}$ is the $2\times 2$ null matrix, and ${\bf 1}$ is the
$2\times 2$ unit matrix.
Substituting this form into $V$, the potential reduces to
\begin{equation}
\label{Vsea}
V=-m_s\,\tr\left(\Sigma_{ss}+\Sigma_{ss}^\dagger\right)
-\frac{1}{2}c_2\left(\tr\left(\Sigma_{ss}+\Sigma_{ss}^\dagger\right)\right)^2\ .
\end{equation}
This is precisely the Sharpe--Singleton potential for the sea sector, which predicts
the Aoki phase for $c_2<0$.  No other non-trivial phase can occur.

\section{Conclusion}
Referring back to the two questions raised toward the end of Sec.~3, the answer to
the first question is no, MAQCD cannot have a mixed phase.  Because the valence
quarks are not part of the dynamics, the phase structure is restricted to that of the
sea sector alone.  In our case, this implies that the only possible non-trivial phase
is an Aoki phase.   The answer to the second question is also no.  In other words,
MAChPT gets it right.  As we have seen, mass inequalities in the underlying theory
restrict the values of LECs in the EFT such as to make regions in the phase diagram
with a mixed phase inaccessible.  We gave an example of such a restriction, and
it is possible that more such constraints on the LECs exist.  These would be uncovered
by a study of the full chiral effective potential for arbitrary $\Sigma_{vacuum}$ by
imposing restrictions following from the Vafa--Witten theorem in the underlying theory.

One might ask whether a similar argument could lead to a restriction on the value
of $c_2$, which, for negative values leads to the existence of the Aoki phase \cite{ShSi}.
The answer is negative, because the argument would have to involve the 
neutral pion, and thus (quark-)disconnected diagrams not captured by the
expression in Eq.~(\ref{start}).   Finally, we note that our conclusions presumably
generalize to other mixed-action theories, such as those with a staggered sea
sector and a chiral valence sector \cite{BBRS}.\\

\leftline{\bf Acknowledgements}
MG thanks the Department of Physics of Humboldt Universit\"at zu Berlin, and YS
thanks the Department of Physics and Astronomy
of San Francisco State University for hospitality.
OB is supported in part by the Deutsche Forschungsgemeinschaft (SFB/TR 09), MG
is supported in part by the US Department of Energy, and YS is supported by the
Israel Science Foundation under grant no. 423/09.


\begin{thebibliography}{99}
\bibitem{BGPQ}  
C.~Bernard and M.~Golterman
  Phys.\ Rev.\  D {\bf 49}, 486 (1994)
  [arXiv:hep-lat/9306005].
  
\bibitem{BRS2}
O.~B\"ar, G.~Rupak and N.~Shoresh,
  Phys.\ Rev.\  D {\bf 70}, 034508 (2004)
  [arXiv:hep-lat/0306021].

\bibitem{BRS1}
O.~B\"ar, G.~Rupak and N.~Shoresh,
  Phys.\ Rev.\  D {\bf 67}, 114505 (2003)
  [arXiv:hep-lat/0210050].
  
\bibitem{GIS}
M.~Golterman, T.~Izubuchi and Y.~Shamir,
  Phys.\ Rev.\  D {\bf 71}, 114508 (2005)
  [arXiv:hep-lat/0504013].

\bibitem{GW}
  P.~H.~Ginsparg and K.~G.~Wilson,
  Phys.\ Rev.\  D {\bf 25}, 2649 (1982).

\bibitem{BGS}
O.~B\"ar, M.~Golterman and Y.~Shamir,
  Phys.\ Rev.\  {\bf D83}, 054501 (2011)
  [arXiv:1012.0987 [hep-lat]].

\bibitem{MGLH}
  M.~Golterman,
  in ``{\it Modern Perspectives in Lattice QCD},'' eds. L.~Lellouch {\it et al.}
  (Les Houches 2009), Oxford
  [arXiv:0912.4042 [hep-lat]].

\bibitem{ShSi}
  S.~R.~Sharpe and R.~L.~Singleton~Jr.,
  Phys.\ Rev.\  D {\bf 58}, 074501 (1998)
  [arXiv:hep-lat/9804028].

\bibitem{Aoki}
S.~Aoki,
  Phys.\ Rev.\  D {\bf 30}, 2653 (1984).

\bibitem{VW}
C.~Vafa and E.~Witten,
  Nucl.\ Phys.\  B {\bf 234}, 173 (1984).

\bibitem{BBRS}
O.~B\"ar, C.~Bernard, G.~Rupak and N.~Shoresh,
  Phys.\ Rev.\  D {\bf 72}, 054502 (2005)
  [arXiv:hep-lat/0503009].

\end{thebibliography}
\end{document}